\documentclass[a4paper,preprint,groupedaddress,showpacs,nofootinbib]{revtex4}

\usepackage{amssymb,amsmath,graphicx}
\usepackage[dvipsnames]{color}

\begin{document}

\title[A Unified Model of Dark Energy]%
      {A Unified Model of Dark Energy Based on \\
       the Mandelstam--Tamm Uncertainty Relation}

\author{Yurii V. Dumin}

\email[]{dumin@yahoo.com, dumin@sai.msu.ru}

\affiliation{
Sternberg Astronomical Institute (GAISh), Lomonosov Moscow State University,\\
Universitetskii prosp.\ 13, Moscow, 119234 Russia}

\affiliation{
Space Research Institute (IKI), Russian Academy of Sciences,\\
Profsoyuznaya str.\ 84/32, Moscow, 117997 Russia}

\begin{abstract}
It is commonly recognized now that Dark Energy (Lambda-term) is of crucial
importance both at the early (inflationary) stage of cosmological evolution
and at the present time.
However, little is known about its nature and origin till now.
In particular, it is still unclear if Lambda-term is a new fundamental
constant or represents just an effective contribution from the underlying
field theory.
Here, we show that a quite promising and universal approach to this problem
might be based on the Mandelstam--Tamm uncertainty relation of quantum
mechanics.
As a result, we get the effective Lambda-term that is important throughout
the entire history of the Universe.
Besides, such an approach requires a substantial reconsideration of some
other cosmological parameters, e.g., the age of the Universe.
\end{abstract}

\pacs{98.80.Bp, 98.80.Cq, 98.80.Es, 95.36.+x}
%
% 98.80.Bp Origin and formation of the Universe
% 98.80.Cq Particle-theory and field-theory models of the early Universe
%          (including cosmic pancakes, cosmic strings, chaotic phenomena,
%          inflationary universe, etc.) 
% 98.80.Es Observational cosmology (including Hubble constant, distance scale,
%          cosmological constant, early Universe, etc)
% 95.36.+x Dark energy

% \received{April 20, 2018}

\maketitle

\section{Introduction}

As is known, the $ \Lambda $-term (or cosmological constant) was introduced
by Einstein immediately during development of General Relativity, but
it was thought for a long time to be insignificant in cosmology.
The situation changed in the early 1980's, when the inflationary models
were suggested to resolve the problem of homogeneity of the early Universe,
the absence of singularity, etc.~\cite{Starobinsky_1980,Linde_1984}.
The crucial ingredient of such models was the stage of exponential expansion
caused by the dominance of the effective $ \Lambda $-term, resulting from
some kind of the underlying theory of elementary particles (such as
supersymmetry, Higgs fields, etc.).

About two decades later, the redshift distribution of supernovae~Ia as well
as the analysis of CMB fluctuations enforced us to conclude that the modern
evolution of the Universe is also governed by the $ \Lambda $-term, but
with much less magnitude than during inflation.
Such $ \Lambda $-term was pictorially called the Dark Energy.

So, the current point of view~\cite{Mortonson_2016} is that
the $ \Lambda $-term plays a decisive role both in the very early and
present-day Universe.
However, its values at these stages are absolutely different and,
therefore, it is unclear if such $ \Lambda $-terms are produced by
the same physical mechanism or they are absolutely different physical
entities?
It is the aim of the present paper to demonstrate that a quite universal
explanation of the time-dependent $ \Lambda $-term, which is significant at
different stages of cosmological evolution, can be given by the
quantum-mechanical uncertainty relation in the Mandelstam--Tamm form.

\section{Theoretical Model}

The standard Heisenberg uncertainty relation
\begin{equation}
\Delta x \, \Delta p \ge \frac{1}{2} \, \hbar
\label{eq:Uncer_rel_xp}
\end{equation}
for variances of the coordinate and momentum (or any other pair of
non-commuting quantities) can be obtained by the straightforward averaging
of the corresponding quantum operators.

The situation is much more subtle for the relation between the energy
and time
\begin{equation}
\Delta E \, \Delta t \ge \frac{1}{2} \, \hbar \: ,
\label{eq:Uncer_rel_Et}
\end{equation}
because the time is an independent parameter in the quantum-mechanical
equations rather than the operator acting in the Hilbert space of
quantum states.
In fact, the inequality~(\ref{eq:Uncer_rel_Et}) was written by Heisenberg
from the dimensionality arguments already in the early days of quantum
mechanics.
However, its first rigorous proof in the framework of a particular model
was given only about two decades later~\cite{Mandelstam_1945} and,
therefore, it is usually called in the modern literature the Mandelstam--Tamm
(rather than Heisenberg) uncertainty relation.
A comprehensive review of the recent works on various interpretations of
inequality~(\ref{eq:Uncer_rel_Et}) can be found in paper~\cite{Deffner_2017}.

Apart from their primary application to the measurement problems,
the uncertainty relations can be efficiently used for estimating
the characteristic parameters of the quantum systems.
(The most well-known example is the evaluation of typical size and energy of
the ground-state hydrogen atom, provided that the equality sign is employed.)
So, the basic idea of our subsequent consideration will be to use
the Mandelstam--Tamm uncertainty relation for getting the quantum-mechanical
vacuum energy, associated with $ \Lambda $-term.

Returning to the cosmological model, we start with the usual
Robertson--Walker metric:
\begin{equation}
d s^2 = c^2 dt^2
  - R^2(t) \bigg[ \frac{\displaystyle dr^2}{\displaystyle 1 - k r^2}
  + r^2 (d {\theta}^2 + {\sin}^2 \theta \, d {\varphi}^2) \bigg] \: ,
\end{equation}
where $ c $~is the speed of light,
$ R $~is the scale factor of the Universe;
$ r $, $ \theta $, and $ \varphi $~are the dimensionless coordinates;
$ k = $ 1, 0, and --1 for the closed, flat, and open three-dimensional
space.

For this metric, the General Relativity equations are reduced to
the Friedmann equation~\cite{Olive_2016}:
\begin{equation}
H^2 \equiv \bigg( \frac{\dot{R}}{R} \bigg) ^{\! 2} =
  \frac{8 \pi G}{3 c^2} \, \rho
  \, - \, k c^2 \frac{1}{\displaystyle R^{\, 2}}
  \, + \, \frac{c^2}{3} \, \Lambda \: ,
\label{eq:Friedmann}
\end{equation}
where $ H $~is the Hubble parameter,
$ G $~is the gravitational constant,
$ \rho $~is the energy density of matter in the Universe, and
dot denotes differentiation with respect to time.

The $ \Lambda $-term can be related to the vacuum energy
density~$ {\rho}_{\rm v} $~\cite{Olive_2016}:
\begin{equation}
\Lambda = \frac{\displaystyle 8 \pi G {\rho}_{\rm v}}{c^4} \, .
\end{equation}
Next, assuming that $ \Delta E = {\rho}_{\rm v} l_{\rm P}^{\, 3} $
is the vacuum energy in the Planck volume
(where $ l_{\rm P}{=}\sqrt{G \hbar / c^3} $),
and $ \Delta t \equiv t $ is the total time of cosmological evolution,
we can apply the Mandelstam--Tamm uncertainty
relation~(\ref{eq:Uncer_rel_Et}) with the equality sign.
This results in
\begin{equation}
\Lambda (t) = \frac{4 \pi}{c \, l_{\rm P}} \, \frac{1}{t} \, .
\label{eq:Lambta_t}
\end{equation}
Finally, substituting expression~(\ref{eq:Lambta_t})
into~(\ref{eq:Friedmann}), we get the basic equation of our
cosmological model:
\begin{equation}
H^2 \equiv \bigg( \frac{\dot{R}}{R} \bigg) ^{\! 2} =
  \frac{8 \pi G}{3 c^2} \, \rho
  \, - \, k c^2 \frac{1}{\displaystyle R^{\, 2}}
  \, + \, \frac{4 \pi}{3 \tau} \, \frac{1}{t} \: ,
\label{eq:Friedmann_modif}
\end{equation}
where $ \tau = l_{\rm P} / c = \sqrt{G \hbar / c^5} $ is the Planck time.

To reveal the most important features of this equation, let us consider
the case when the Universe is spatially flat ($ k{=}0 $), and the energy
density of matter is ignorable ($ \rho \approx 0 $).
Then, formula~(\ref{eq:Friedmann_modif}) is simplified to
\begin{equation}
H^2 \equiv \bigg( \frac{\dot{R}}{R} \bigg) ^{\! 2} =
  \frac{4 \pi}{3 \tau} \, \frac{1}{t} \: ,
\label{eq:Friedmann_simplif}
\end{equation}
which can be trivially integrated and results in
\begin{equation}
R(t) = R^* \exp \bigg[ \sqrt{\frac{16 \pi}{3}} \,
  \sqrt{\frac{t}{\tau}} \, \bigg] \, ,
\label{eq:R_t}
\end{equation}
where the integration constant was chosen so that $ R(0) = R^* $,
and we consider only the solution increasing with time.

Finally, let us note that according to formula~(\ref{eq:Friedmann_modif})
\begin{equation}
T = \frac{4 \pi}{3} \, \frac{1}{\tau H_0} \, \frac{1}{H_0} \: ,
\label{eq:age_Universe}
\end{equation}
where $ T $~is the age of the Universe, and
$ H_0 $~is the present-day value of the Hubble parameter.

\section{Discussion and Conclusions}

\begin{enumerate}

\item
The proposed cosmological model provides a natural explanation for
the existence of Dark Energy (effective $ \Lambda $-term) throughout
the entire life of the Universe.

\item
As distinct from the ``standard'' cosmology, where scale factor~$ R(t) $
evolves either exponentially (when the Dark Energy is dominant) or by
a power law (when ordinary matter dominates), in our model $ R(t) $ evolves
by the same universal law~(\ref{eq:R_t}), which is much slower than a pure
exponential expansion but much faster than any power-like dependence.

\item
While in the standard cosmology the Hubble parameter either remains constant
with time (when evolution is determined by the $ \Lambda $-term) or decays
as~$ \alpha / t $ (where $ \alpha = 1/2 $ when radiation dominates, and
$ \alpha = 2/3 $ when a non-relativistic matter dominates),
our equation~(\ref{eq:Friedmann_simplif}) predicts the inverse square-root
dependence $ H(t) \propto 1 / \sqrt{t} $, i.e., again some intermediate case
between the two extremal situations.

\item
Taking into account that age of the Universe in the standard cosmology is,
roughly speaking, inversely proportional to the present-day value of
the Hubble parameter, $ T^* \approx 1 / H_0 $, we see that our
relation~(\ref{eq:age_Universe}) predicts that
$ \, T \approx (T^* \! / \tau) \, T^* $.
Since $ T^* \approx 4{\cdot}10^{17} $\,s, and
$ \, \tau = 5{\cdot}10^{-44} $\,s,
this age in our model will be increased by
$ (T^* \! / \tau) \approx 10^{61} $ times.
In other words, the Universe becomes ``quasi-perpetual''.
However, this anomalous lifetime should not be a fatal failure of
the model: Really, as far as we know, the most of problematic issues in
the modern cosmology are caused just by the insufficient lifetime of
the Universe (so that some astronomical objects do not have sufficient
time to be formed).
On the other hand, it is difficult to say without a detailed analysis if
there will be some crucial obstacles in the case of the anomalously long
lifetime.

\item
It will be desirable, of course, to consider more general solutions of
the equation~(\ref{eq:Friedmann_modif}), involving a few matter components.
Besides, due to the substantially different temporal evolution of
the Universe~(\ref{eq:R_t}) as compared to the ``standard'' model,
the processes of nucleosynthesis, cosmological structure formation, etc.\
should be carefully recalculated.

\item
At last, one of hot topics of cosmology in the recent years is a systematic
discrepancy in the values of the present-day Hubble parameter~$ H_0 $ derived
by the various methods (namely, by the employment of supernovae~Ia and
Cepheids as the ``standard candles'', on the one hand, and by the analysis of
CMB spectrum, on the other hand)~\cite{Ryden_2017,Freedman_2017}.
Since in the second case the resulting value of~$ H_0 $ substantially depends
on the expansion history of the Universe~$ R(t) $, the most of the recent
attempts to resolve the above-mentioned discrepancy were based on
the empirical modifications of the equation of state of the Dark
Energy~\cite{Huang_2016,Liu_2016,Di_Valentino_2017,Zhao_2017}.
In this sense, the substantially modified temporal dependence~(\ref{eq:R_t}),
naturally following from our model, might be an additional option for
such attempts.

\end{enumerate}

\bigskip

I am grateful to A.~Starobinsky for the occasional consultations on various
problems of cosmology.


\begin{thebibliography}{99}

\bibitem{Starobinsky_1980}
A.A.~Starobinsky,
``A New Type of Isotropic Cosmological Models without Singularity'',
\textit{Phys.\ Lett.}
{\bf 91B} 99 (1980).

\bibitem{Linde_1984}
A.D.~Linde,
``The Inflationary Universe'',
\textit{Rep.\ Prog.\ Phys.}
{\bf 47} 925 (1984).

\bibitem{Mortonson_2016}
M.J.~Mortonson, D.H.~Weinberg, and M.~White,
``Dark Energy'',
in: C.~Patrignani, et al.\ (Particle Data Group),
\textit{Review of Particle Physics},
\textit{Chinese Physics C} {\bf 40} 100001 (2016), p.402.

\bibitem{Mandelstam_1945}
L.I.~Mandelstam and I.E.~Tamm,
``The Energy--Time Uncertainty Relation in Non-relativistic
Quantum Mechanics'',
\textit{Izv.\ Akad.\ Nauk SSSR (Ser.\ Fiz.)}
{\bf 9} 122 (1945, in Russian).

\bibitem{Deffner_2017}
S.~Deffner and S.~Campbell,
``Quantum Speed Limits: from Heisenberg's Uncertainty Principle
to Optimal Quantum Control'',
\textit{J.\ Phys.\ A: Math.\ Theor.}
{\bf 50} 453001 (2017).

\bibitem{Olive_2016}
K.A.~Olive and J.A.~Peacock,
``Big-Bang Cosmology'',
in: C.~Patrignani, et al.\ (Particle Data Group),
\textit{Review of Particle Physics},
\textit{Chinese Physics C} {\bf 40} 100001 (2016), p.355.

\bibitem{Ryden_2017}
B.~Ryden,
``A Constant Conflict'',
\textit{Nature Physics} {\bf 13} 314 (2017).

\bibitem{Freedman_2017}
W.L.~Freedman,
``Cosmology at a Crossroads'',
\textit{Nature Astron.} {\bf 1} 0121 (2017).

\bibitem{Huang_2016}
Q.-G.~Huang and K.~Wang,
``How the Dark Energy Can Reconcile Planck with Local Determination of
the Hubble Constant'',
\textit{Eur.\ Phys.\ J.\ C} {\bf 76} 506 (2016).

\bibitem{Liu_2016}
Z.-E.~Liu, H.-R.~Yu, T.-J.~Zhang, and Y.-K.~Tang,
``Direct Reconstruction of Dynamical Dark Energy from Observational
Hubble Parameter Data'',
\textit{Phys.\ Dark Univ.} {\bf 14} 21 (2016).

\bibitem{Di_Valentino_2017}
E.~Di~Valentino,
``Crack in the Cosmological Paradigm'',
\textit{Nature Astron.} {\bf 1} 569 (2017).

\bibitem{Zhao_2017}
G.-B.~Zhao, M.~Raveri, L.~Pogosian, et al.,
``Dynamical Dark Energy in Light of the Latest Observations'',
\textit{Nature Astron.} {\bf 1} 627 (2017).

\end{thebibliography}
\end{document}